\documentclass{aa}

\usepackage{color}

\usepackage{amsmath}
\usepackage{natbib}
\usepackage{graphicx}
\usepackage{makecell}
\usepackage{color}
\begin{document}

   \title{Spectropolarimetric NLTE inversion code SNAPI}


   \author{I. Mili\'{c}\inst{1}
   \and
   M. van Noort\inst{1}}
   
   \institute{Max-Planck-Institut f\"{u}r Sonnersystemforschung, Justus-von-Liebig-Weg 3, 37077 G\"{o}ttingen, Germany\\
   \email{milic@mps.mpg.de; vannoort@mps.mpg.de}}
         
   \date{}
   \titlerunning{NLTE inversion code SNAPI}
   \authorrunning{I. Mili\'{c} \& M. van Noort}


  \abstract
   {Inversion codes are computer programs that fit a model atmosphere to the observed Stokes spectra, thus retrieving the relevant atmospheric parameters. The rising interest in the solar chromosphere, where spectral lines are formed by scattering, requires developing, testing, and comparing new  non-local thermal equilibrium (NLTE) inversion codes.}  
   {We present a new NLTE inversion code that is based on the analytical computation of the response functions. We named the code SNAPI, which is short for spectropolarimetic NLTE analytically powered inversion.}
   {SNAPI inverts full Stokes spectrum in order to obtain a depth-dependent stratification of the temperature, velocity, and the magnetic field vector. It is based on the so-called node approach, where atmospheric parameters are free to vary in several fixed points in the atmosphere, and are assumed to behave as splines in between. We describe the inversion approach in general and the specific choices we have made in the implementation.}
   {We test the performance on one academic problem and on two interesting NLTE examples, the Ca\,II\,8542 and Na\,I\,D spectral lines. The code is found to have excellent convergence properties and outperforms a finite-difference based code in this specific implementation by at least a factor of three. We invert synthetic observations of Na lines from a small part of a simulated solar atmosphere and conclude that the Na lines reliably retrieve the magnetic field and velocity in the range $-3<\log \tau < -0.5$}. 
 

   \maketitle


\section{Introduction}


Current modern solar telescopes provide us with high angular ($\approx0.1''$) and spectral ($\lambda/\Delta\lambda \geq 10^5$) resolution spectropolarimetric observations, with a high signal-to-noise ratio ($\approx 10^3$). Such high-quality data are especially valuable because they allow resolving the shapes of the spectral lines
that are formed in the solar atmosphere. Across spectral lines, the atmospheric opacity changes dramatically in a narrow range of wavelengths, causing the radiation we receive to escape from very different depths in the solar atmosphere. Since the shape and wavelength of atomic line profiles are sensitive to the velocity and magnetic field properties, the shape of the resulting spectral line encodes information about the physical conditions (temperature $T$, velocity $\vec{v,}$ and the magnetic field $\vec{B}$) as a function of depth in the solar atmosphere. 

The method of choice to recover these depth-dependent atmospheric properties is spectropolarimetric inversion. Inversion is the process of fitting a model atmosphere to an observed Stokes spectrum in a given pixel. By inverting the observations, we obtain a model atmosphere for each of the observed pixels, and thus a pseudo-3D structure of the observed patch on the solar surface. We use ``pseudo'' here because the spectra contain almost exclusively information on the stratification of physical quantities in optical depth, and almost no information about the geometrical depth.

Inversion codes allow us to proceed from the space of Stokes spectra $\vec{I}(x,y,\lambda)$, where $\vec{I}=(I,Q,U,V)$, to the space of physical parameters $\vec{\Theta}(x,y,\tau)$, where $\vec{\Theta} = (T,\vec{v},\vec{B})$ and $\tau$ is the optical depth in the continuum along the vertical. The main difficulty in the inversion process is the nonlinear relationship between the atmospheric parameters we wish to infer and the observed Stokes vector. The atmospheric properties determine the opacity and emissivity in the atmosphere, which through the radiative transfer equation in turn determine the emerging polarized intensity. In practice, opacity and emissivity are calculated numerically, as is the solution of the radiative transfer equation. This makes the forward problem (calculation of the Stokes parameters) numerically demanding. 

The past several decades have seen the development of inversion codes with increasing levels of complexity. At first, the Milne-Eddington (ME) assumption was used \citep{Auer77,Egidio82}, which is rather crude, but results in an analytical relationship between the model parameters and the observables, making the inversion process relatively fast. However, the ME approach assumes that the velocity and the magnetic field are independent of depth. This is a weak assumption, since the asymmetries in the Stokes profiles in spatially resolved observations clearly indicate the presence of velocity and magnetic field gradients in the real solar atmosphere. 

Although the SIR \citep{SIR} and SPINOR \citep{SPINOR} inversion codes account for a depth dependence of the relevant physical parameters and allow us to obtain fully depth-stratified atmospheres from the observed Stokes spectra, these codes assume that the matter and radiation are in so-called local thermodynamic equilibrium (LTE), which is a valid assumption only for lines formed in the photosphere, where collisions dominate.

To study the upper layers of the solar atmosphere, the spectra must also contain spectral lines that are formed there. In these layers, the density is so low that collisions no longer dominate and the assumption of LTE no longer holds, a condition often referred to as non-LTE (NLTE), a description of what it is not, rather than what it is. If the atmosphere properties can be considered independent of time, an equilibrium can be assumed to exist between excitation and de-excitation of atoms due to collisions, and due to radiative transitions. In such a statistical equilibrium state, the opacity and emissivity that determine the radiation field are not only determined by the atmospheric properties, but also by the radiation field itself. The dependence of the radiation field on itself requires iterative methods to be solved, making the forward solution of the NLTE problem several orders of magnitude more numerically demanding than if LTE can be assumed. Moreover, the iterative solution of the transfer equation makes it more difficult to predict the change in the spectrum given a change in the atmospheric properties, so that finite differences are traditionally used, at a considerably additional numerical expense. As a consequence, only two inversion codes that can be used for NLTE lines have been described in the literature: NICOLE by \citet{NICOLE}, and STiC by \citet{STIC}.

In this manuscript, we describe a new NLTE inversion code, named SNAPI, which is short for spectropolarimetric NLTE analytically powered inversion, which can make use of the analytical response functions described in  \citet{MMvN17}. We start by briefly outlining the NLTE problem and our recent work on response functions. We then describe the code and discuss the particular atmospheric parameterization we choose (the so-called node approach), the least-squares minimization process used to fit these parameters, and how the analytical response functions can be used to calculate the derivatives needed for this minimization. Finally, we test the code on three problems of increasing complexity and analyze its performance. 

\section{NLTE problem and response functions}

We consider a single, 1D model atmosphere where physical parameters can vary arbitrarily with height. The atmosphere is discretized, and thus all the equations are to be solved numerically. The aim is to calculate the emergent Stokes spectrum $\vec{I,}$ where $\vec{I}=(I,Q,U,V)$ is the so-called Stokes vector that describes the polarization state of the light. $\vec{I}$ is the solution of the polarized radiative transfer equation \citep[e.g.][]{dtibook}
\begin{equation}
\cos \theta \frac{d\vec{I}(z,\theta,\lambda)}{dz} = - \hat{K}(z,\theta,\lambda)\vec{I}(z,\theta,\lambda) + \vec{j}(z,\theta,\lambda),
\label{polRTE}
\end{equation}
where $z$ is the vertical coordinate, $\theta$ is the angle of the propagation of the light with respect to $z$, and $\lambda$ is the wavelength. $\hat{K}$ (a $4\times4$ matrix) and $\vec{j}$ (a $4$-vector) are {the propagation matrix and the emission vector, respectively \citep[for their exact form, see, e.g.,][]{dtibook,LL04}. It is customary to replace the geometrical height $z$ with the
optical depth in the continuum, $\tau$:
\begin{equation}
d\tau = - \chi(\lambda_0,z) dz,
\end{equation}
where $\chi$ is the scalar (i.e., unpolarized) absorption coefficient. In inversions, $\tau$ is used as an independent coordinate, and we aim to infer the stratification of the physical parameters on the $\tau$ grid. The reference wavelength $\lambda_0$ is usually chosen to be $5000\,\rm{\AA}$.


Even though the assumption of LTE is strictly speaking never valid, as the assumption of equilibrium implies that no radiation can actually escape, it is fairly accurate deep in the atmosphere, since the radiative losses there are insignificant compared to the collisional rates, so that the atomic level populations are described adequately by the Saha-Boltzmann equations \citep[see,
e.g.,][]{SAbook2014}. Despite the nonlinear character of these equations, they depend only on local quantities, so that for a given hydrodynamic state, the atomic level populations can be directly calculated.

In the low-density conditions that prevail in the solar chromosphere, however, the radiative rates can be considerably higher than the collisional rates, and the radiative losses cannot be neglected. In order to obtain the atomic level populations in these conditions, we must resort to solving the statistical equilibrium equations at each point in the atmosphere:
\begin{equation}
\sum_j (n_{j} T_{ji} - n_{i} T_{ij}) = 0,
\label{SSE}
\end{equation}
where the total transition rate from level $i$ to level $j$ is given by $T_{ij} = C_{ij} + R_{ij}$, where $C_{ij}$  and $R_{ij}$ are the collisional and radiative  transition rates, respectively, and $n_i$ is the number density of all atoms in state $i$. For spectral line transitions, the radiative rates are of the form
\begin{equation}
R_{ij} = A_{ij} + B_{ij} \int_0^{\pi} \int_{-\infty}^{\infty} I(\theta,\lambda) \phi_{ij}(\theta,\lambda) d\lambda \sin\theta d\theta,
\label{rate}
\end{equation}
where $A_{ij}$ and $B_{ij}$ are the Einstein coefficients for spontaneous and stimulated emission, respectively, for $i>j$, or zero and the Einstein absorption coefficient, for $i<j$, respectively. We assume that the frequency dependence of the absorption and emission probabilities are identical and described by the profile function $\phi_{ij}(\theta,\lambda)$, and that the frequencies of any photons that are absorbed and then re-emitted are uncorrelated.
This is known generally as complete frequency redistribution (CRD), which is a good approximation for the transitions considered in this paper. In the case of bound-free processes, the equations have a similar form.  

The integral on the right-hand side is sometimes referred to as the scattering integral, and it contains the intensity of the radiation field in the solar atmosphere. To first order, the intensity $I(\theta,\lambda)$ can be calculated from the solution of the scalar (i.e., unpolarized) radiative transfer equation
\begin{equation}
\cos \theta \frac{dI(\tau,\theta,\lambda)}{d\tau_{\lambda}} = I(\tau,\theta,\lambda) - S(\tau,\theta,\lambda),
\label{scalarRTE}
\end{equation}
where $d\tau_{\lambda}=d\tau \chi(\lambda)/\chi_c$ is the monochromatic optical depth, and $S$ is the ratio of the emissivity to the opacity, commonly referred to as the source function. Since $S$ and $\tau_{\lambda}$ both depend on the atmospheric properties but also on the level populations, clearly, Eqs.
(\ref{SSE}) and (\ref{scalarRTE}) form a coupled system that needs to be solved self-consistently. The nonlinearity of the scattering integral, combined with the nonlocal character of the radiation field, ensures that an analytical solution cannot generally be found. The solution must therefore be obtained iteratively, and a number of methods have been described in the literature \citep[see, e.g.,][]{Hubeny03,SAbook2014}. For SNAPI we have opted for the tried-and-tested method MALI (multi-level accelerated lambda iteration) developed by \citet{RH1}. 

\subsection{Analytical calculation of the response functions}
\label{RFS}

After iteratively obtaining the level populations $n_i$, we are able to numerically solve Eq.\,\ref{polRTE} and obtain the emergent Stokes spectrum. Since it was produced by an arbitrary atmospheric structure, the calculated spectrum will generally not match the observed one. The aim of the inversion process is to modify the atmospheric structure until the agreement between the synthetic and the observed Stokes spectra is satisfactory. This task is performed by a majority of similar inversion codes using a derivative-based minimization procedure. The derivatives required for this type of minimization require knowledge of the response of the synthetic spectrum to a given perturbation in the atmospheric structure, the so-called response functions,
\begin{equation}
\mathcal{R}(q,\tau_d,\lambda) = \frac{\partial\hat{I}^+(\lambda)}{\partial q(\tau_d)},
\end{equation}
where the superscript $+$ indicates the outgoing direction of the radiation, and $\tau_d$ denotes the continuum optical depth at the location $z_d$ in the atmosphere.

Response functions can conveniently be obtained using a finite-difference approximation, that is, by perturbing the atmospheric structure and measuring the resulting spectral change. Although this method is robust, very simple to implement, and accurately returns the true response of the sytnhesized spectra as calculated by the radiative transfer solver (provided the appropriate step size is used), it is a slow method if the spectral calculations are time consuming.

Fortunately, the response functions for the emergent intensity follow from the response functions for polarized emissivity and opacity, according to Eq.\,\ref{polRTE}, which in turn depend on the response functions of the level populations. Since in LTE these level populations are given by the  Saha and Boltzmann equations, the response functions $\mathcal{R}(q,\tau_d,\lambda)$ can be readily calculated analytically.

In NLTE, however, the level populations are obtained numerically, and an exact analytical expression cannot generally be obtained. The obvious alternative, a finite-difference based numerical approach of the level populations, is numerically highly demanding, since the NLTE problem needs to be solved many times, and it
does not provide an effective alternative to the direct finite-difference calculation of the spectral response. 

Analytical expansion of all the terms in the statistical equilibrium equation (Eq.\,\ref{SSE}), however, reveals that it is possible to obtain the response functions for the level populations in NLTE in a more direct way. We refer to the derivation and the tests given in \citet{MMvN17}, where we also discuss response functions in much greater detail. 

Although we stress that no analytical solution of the NLTE problem is possible in the case of an arbitrarily stratified atmosphere, and therefore no truly analytic response functions can be found, we refer to the semi-analytical method described in this paper as the ``analytical'' method for convenience. SNAPI uses this analytical method for computing the response functions by default, and the finite-difference approach is only used for comparison and verification.

Armed with the response functions, we are able to modify the atmospheric structure in a way that reduces the difference between the synthesized and observed spectra, but the details of this modification are far from trivial and are at the core of the inversion code.

\section{Atmospheric parameterization and inversion procedure}

The response functions yield partial derivatives of the observable (the emergent Stokes vector) with respect to the values of each physical parameter at each depth point in the atmosphere. For completeness, we list the physical parameters and the way these influence the observable:
\begin{itemize}
 \item temperature: collisional rates, line broadening 
 \item particle density: collisional rates, total number of absorbers/emitters
 \item microturbulent velocity: line broadening
 \item line-of-sight velocity: line absorption/emission profile shifts 
 \item magnetic field vector ($B,\theta_B,\phi_B$): polarized absorption matrix. 
\end{itemize}
Since the density can be eliminated from this set by assuming that the atmosphere is in hydrostatic equilibrium \citep[see, e.g.,][]{BV89}, it is sufficient to characterize the atmosphere with a total of six independent physical parameters in each depth point. Since a typical numerical discretization rerequires $\mathcal{O}(10^2)$ points to accurately represent the atmosphere, we need to fit $\mathcal{O}(10^2-10^3)$ variables in total to the observed data. 

Unfortunately, the nonlocal character of the radiative transfer suppresses the response of the emerging radiation to small-scale variations in the atmosphere, so that fitting such a complicated model to the observed data is usually not constrained by the data (typically consisting of $\mathcal{O}(10^1-10^2)$ wavelength points in each of the Stokes parameters). If each depth point were to be treated as an independent parameter, the inversion problem would become severely degenerate (ill-posed), and deployment of a derivative-based iterative fitting method would result in unrealistic, erratic corrections to the parameters. To solve this problem, most inversion codes regularize the inversion problem in one of two ways:
\begin{enumerate}
\item Parameterization of corrections: Given a starting atmosphere of arbitrary complexity, we can require the corrections to the atmospheric structure to be sparse in a particular space. Enforcing sparsity is a powerful tool that can help in many areas of inference \citep[e.g.,][]{SparseInv}. Although many approaches are possible, the most frequently used is the so-called node approach, where the atmosphere is independently modified in several points, while the changes in between these points are assumed to behave according to a polynomial. The first application of this method was in the SIR inversion code \citep[][]{SIR,dtibook}. 
\item Parameterization of the atmosphere: This concept is very similar to the node concept from the practical point of view, but rather different from the interpretation point of view. Nodes are used, but this time, to describe the stratification of the atmosphere itself. This means that the general dependencies of physical parameters with depth are assumed to behave like polynomials in between the nodes. This seems very similar to the regularization approach, but in this case, values at the nodes are actually atmospheric properties. The SPINOR code \citep{SPINOR} uses this kind of atmospheric parameterization.
\end{enumerate}

For SNAPI we have currently opted for the second approach. Our main motivation is full control over the model parameters and relative simplicity of the resulting atmospheres (see Fig.\,\ref{node_example} for an example of node-based parametrization of the atmosphere). 

\subsection{Fitting}

After the parameterization outlined in the previous section, we are typically left with a strongly simplified atmosphere to fit to the observations. Fitting spectropolarimetic data usually proceeds with a minimization of a so-called merit function, commonly denoted by $\chi^2$ (not to be confused with the unpolarized opacity $\chi$), defined as the weighted sum of the squared differences between the data and the synthesized spectrum: \begin{equation}
\chi^2 = \sum_s \sum_l w_{s,l} \frac{(I^{\rm calc}_{s,l}-I^{\rm obs}_{s,l})^2}{\sigma^2}
,\end{equation}
where the indices $s$ and $l$ correspond to the different Stokes components and wavelengths, respectively, and $calc$ and $obs$ refer to calculated and observed quantities. $\sigma$ is the standard deviation of the noise in the continuum, which is assumed to be additive and Gaussian. The $w_{s,l}$ are additional weighting factors that can be used to account for any wavelength dependence of the noise, or a different weighting of the four Stokes parameters (e.g., to account for different polarimetric efficiencies, or to satisfy the specific desire of the user to fit certain Stokes components better than others). Additionally, since the noise in the observations is almost universally underestimated and/or poorly understood, these weighting coefficients can be used to ensure that the residuals between the observations and the best-fit solution do indeed describe a Gaussian distribution. Since the calculated Stokes spectra, $I^{calc}$, and thus $\chi^2$, depend on the atmospheric parameters, we can minimize the merit function by adjusting the atmospheric parameters.

Given a suitable parameterization, we define the transformation\begin{equation}
 \vec{\alpha} \rightarrow \vec{\Theta}(\tau)
 \label{eq:transform}
,\end{equation}
where $\vec{\alpha}$ denotes the set of fit parameters (node values) and $\vec{\Theta}$ represents the complete set of depth-dependent physical parameters. We formally write the merit function in terms of the fit parameters $\vec{\alpha}$
$$
\chi^2(\vec{\Theta}) \rightarrow \chi^2(\vec{\Theta}(\vec{\alpha})),
$$
or simply, $\chi^2(\vec{\alpha})$. To find $\vec{\alpha}$ for which $\chi^2(\vec{\alpha})$ is minimal, the method of choice is the Levenberg-Marquardt algorithm \citep{LM_L,LM_M}. 

Starting from an initial guess for the model $\vec{\alpha}$, we compute the forward solution $\vec{I}(\vec{\alpha})$, and evaluate the residuals:
\begin{equation}
r_{s,l} = I^{\rm calc}_{s,l}-I^{\rm obs}_{s,l}.
\end{equation}
We then calculate the partial derivatives of the calculated spectrum to the model parameters (node values),\begin{equation}
 J_{s,l,i} = \frac{\partial I_{s,l}}{\partial \alpha_i}
,\end{equation}
using the response functions $\mathcal{R}(q,\tau_d,\lambda)$ and the transformation defined in Eq. (\ref{eq:transform}).
We then obtain the correction to the current model, $\delta\vec{\alpha}$, by solving the linear equation
\begin{equation}
\hat{H} \delta \vec{\alpha} = \hat{J}^{\rm T} \vec{r},\end{equation}
where $\hat{H}$ is the so-called Hessian matrix,\begin{equation}
\hat{H} = \hat{J}^{\rm T} \hat{J} + \lambda \,{\bf diag}\left[{\hat{J}^{\rm T} \hat{J}}\right].
\label{hessian}
\end{equation}
The constant $\lambda$ is the Marquardt constant, which controls the behavior of the solution, and must be adjusted according to the level of linearity of the merit function. When the merit function is sufficiently linear and $\lambda$ is small, the method reduces to the Newton-Raphson method and converges rapidly. However, when the merit function is sufficiently nonlinear, the error in the correction $\delta\vec{\alpha}$ is so large that addition to $\delta\vec{\alpha}_n$ even increases the value of the merit function, which results in a diverging solution. In this situation, the Marquardt constant is increased, the effect of which is both to reduce the size of the correction and to point it increasingly in the direction of the gradient of the merit function, given by the diagonal of $\hat{J}^{\rm T} \hat{J}$. In the limit of large $\lambda$, the method reduces to the steepest-descent method, for which the solution is guaranteed to converge, albeit slowly. Continuous optimization of the Marquardt constant is therefore an integral part of the Levenberg-Marquardt algorithm, which can have a significant effect on the convergence properties.

Calculation of the Hessian also allows us to estimate the uncertainties of the inferred parameters. We have implemented a simple method described in \citet[][see Eq.\,11.4]{dtibook}. It is important to keep in mind that these uncertainties only describe the uncertainty of each of the inferred parameters relative to that of the other ones. That is, we see that some node values are better constrained than others, but we do not find actual credibility intervals, nor the covariances between the node values. The only methods that can reliably estimate model parameter uncertainties and their covariances are sampling methods \citep[e.g., Markov chain Monte Carlo, see][]{Hogg}, but they are out of the question here
because of the prohibitively expensive forward calculation. An alternative, used by some researchers, is to run the inversion from different initial models and then estimate the uncertainties from the scatter of the best-fit solution. We do not argue in favor of this method, because a large scatter might be caused by local minima, whereas a small scatter might be due to the poor sampling of the $\chi^2$ hypersurface by the initial models. In addition, this approach is not feasible for the inversion of huge sets of data. A rigorous treatment of the inferred model uncertainties is definitely lacking in the field.

\subsection{Response functions to model parameters}
Since the atmosphere was parametrized using nodes, we now need the response functions to the model parameters $\alpha_i$. These can be calculated using the depth-dependent response functions $\mathcal{R}(q,\tau_d,\lambda)$ from Sect. \ref{RFS},
\begin{equation}
\frac{\partial \vec{I}}{\partial \alpha_i} = \sum_d \frac{\partial \vec{I}}{\partial q_d} \frac{\partial{q_d}}{\partial \alpha_i} = \sum_d \mathcal{R}(q, z_d) \frac{\partial{q_d}}{\partial \alpha_i}
\label{param_responses}
,\end{equation}
where $q_d$ stands for the value of the atmospheric property $q$ at depth point $d$. The partial derivative $\frac{\partial{q_d}}{\partial \alpha_i}$ follows from the specific choice of the atmospheric parameterization, that is, from the transformation Eq. (\ref{eq:transform}).

SNAPI assumes that the values between the nodes behave as second-order Bezier polynomials, which are particularly suitable for ensuring that the interpolated quantities will behave monotonically between the nodes. Extrapolation outside the domain covered by the nodes is not always straightforward, and different strategies need to be used for different atmospheric properties (see Table\,\ref{nodes_table}). The $\frac{\partial{q_d}}{\partial \alpha_i}$ follow from the specific expressions for the interpolation and extrapolation. An exception is the derivative of the density with respect to the temperature nodes, which is calculated numerically, since the change in temperature changes the density everywhere through hydrostatic equilibrium.

\begin{figure}
\includegraphics[width = 0.5\textwidth]{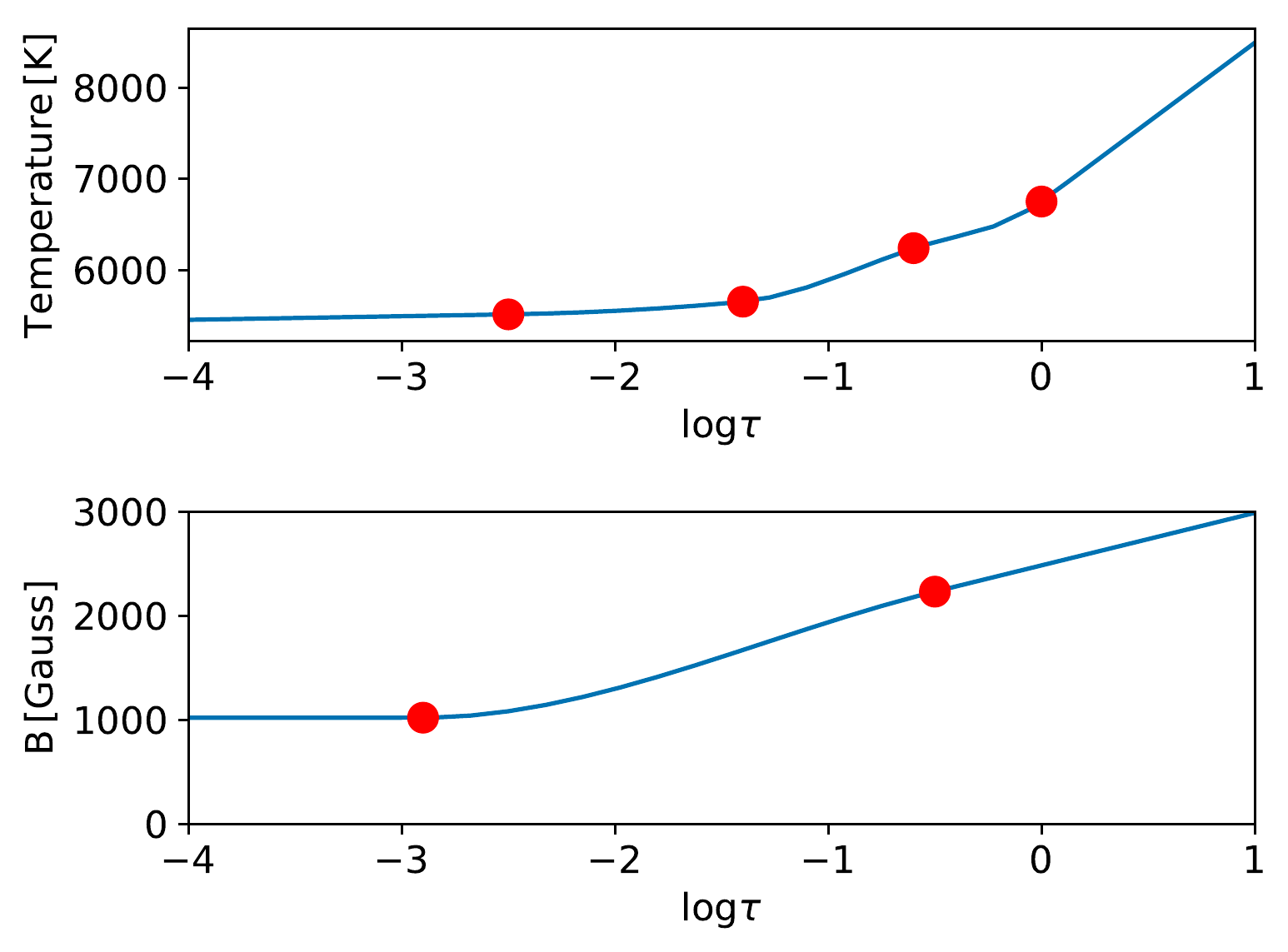}
\caption{Example of nodes used to parametrize an atmosphere. The temperature is described with four nodes (red circles) and the magnetic field with two. Blue lines trace the parameter values on the dense tau grid that is used for the NLTE solution and spectrum synthesis.}
\label{node_example}
\end{figure}

\begin{table}
\caption{Interpolation and extrapolation used to construct the atmosphere from given nodes. The temperature at the bottom of the atmosphere is extrapolated linearly so that its temperature gradient is parallel to the temperature gradient of the semi-empirical FALC model atmosphere \citep{FALC}.} 
\label{nodes_table}      
\centering                                      
\begin{tabular}{c c c c}
\hline
Parameter & Interpolation & Top & Bottom \\
\hline                                  
    $T$ & Bezier & linear & linear$^*$  \\     
    $v$ & Bezier & constant & linear \\
    $B$ & Bezier & constant & linear \\
    $\theta_B,\phi_B$ & Bezier & constant & constant \\
\hline                                             
\end{tabular}
\end{table}

In practice, it is not advantageous to use Eq.\,\ref{param_responses} directly, but to use the same approach as above to obtain the responses of the absorption matrix and emissivity vector to $\vec{\alpha}$ instead:
\begin{eqnarray}
\frac{\partial \hat{K}}{\partial \alpha_i} = \sum_d \frac{\partial \hat{K}}{\partial q_d} \frac{\partial{q_d}}{\partial \alpha_i}, \nonumber \\
\frac{\partial \vec{j}}{\partial \alpha_i} = \sum_d \frac{\partial \vec{j}}{\partial q_d} \frac{\partial{q_d}}{\partial \alpha_i}.
\label{op_em_responses_p}
\end{eqnarray}
The final step is the calculation of $\frac{\partial \vec{I}}{\partial \alpha_i}$ from the responses of $\hat{K}$ and $\vec{j}$ to model parameters. The simplest solution is to do this numerically. It requires two formal solutions per parameter, which takes negligible computational time with respect to the NLTE calculation of atomic populations and their responses. In the appendix of \citet{MMvN17}, we proposed another method for the propagation of the opacity and emissivity responses. We take the explicit, analytical derivatives of the whole process of the formal solution, equation by equation. The process is cumbersome (especially for the polarized case), but straightforward, and it yields responses that agree very well with numerically calculated responses. In SNAPI, we have the option of using either of these approaches.

Finally, after obtaining $\frac{\partial \vec{I}}{\partial \alpha_i}$, we can propose corrections to the current model of the atmosphere. The whole inversion process, starting from the initial atmospheric model specified by giving node positions and values, is depicted in Fig.\,\ref{inversion_flowchart}.

\begin{figure}
\includegraphics[width = 0.5\textwidth]{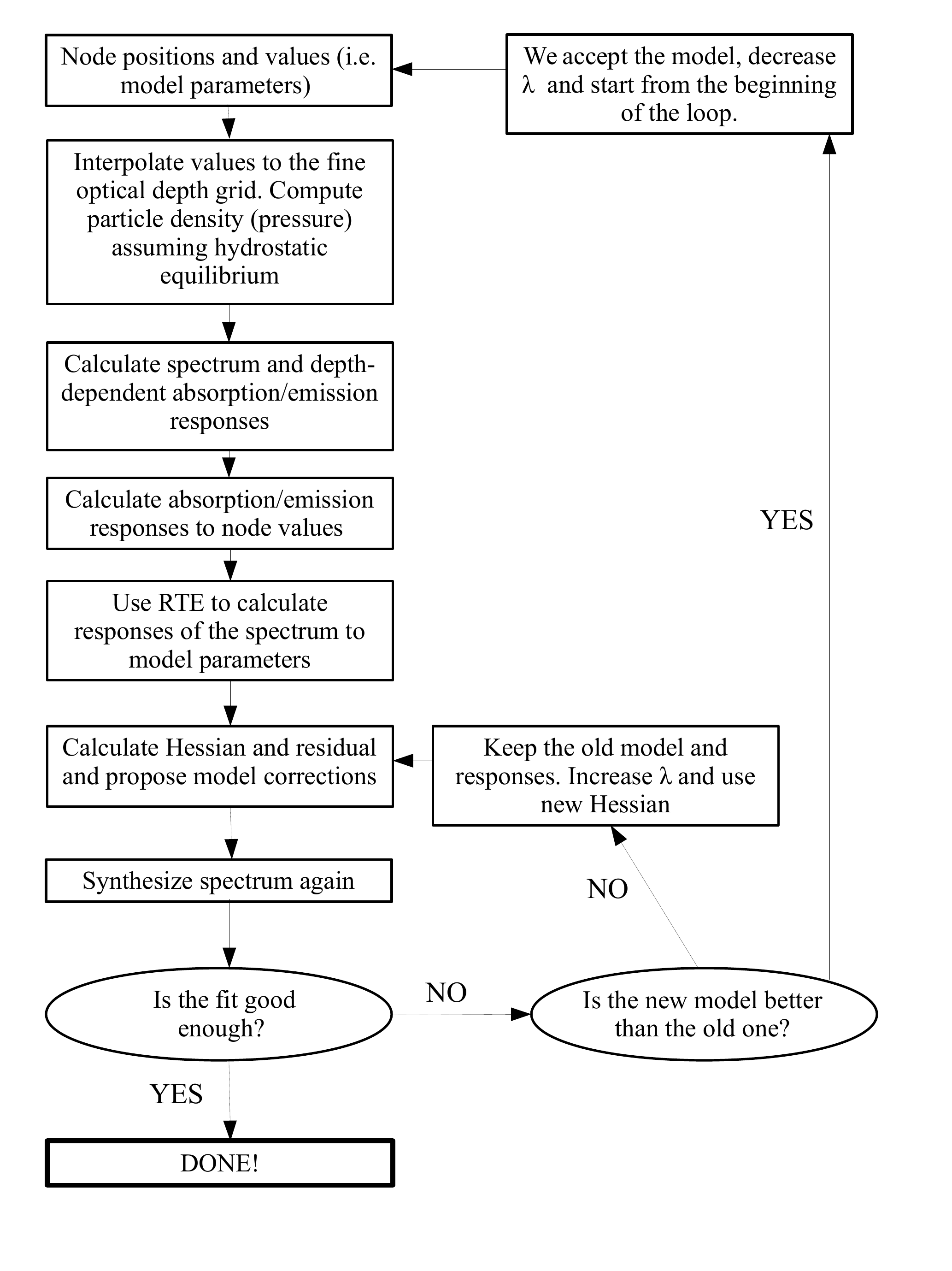}
\caption{Flowchart of the inversion procedure in SNAPI.}
\label{inversion_flowchart}
\end{figure}

\subsection{Implementation notes}

SNAPI is written in C++, using object-oriented programming and inheritance. The model (the node positions and values), atmosphere, atoms, and molecules are treated as objects, while spectrum synthesis, response function calculation, construction of the atmosphere from given nodes, and correction of the model (or the atmosphere) are methods of the appropriate objects. Parallelization is achieved using the master-slave concept, where the master maintains a queue of jobs to be executed. For each job that reaches the head of the queue, the master loads the configuration and the data files, and distributes the work among the slaves it has available, which can be anywhere in the accessible network. Slaves can connect and disconnect randomly, and work on the inversion or spectrum synthesis (whichever mode is chosen) on a pixel-by-pixel basis. When a slave has completed a particular task, the fitted node values, the appropriately inter- or extrapolated atmosphere, and the fitted spectra are sent to the master. When the job is completed, the master stores the inverted node, atmosphere, and spectral hypercubes to the disk, and begins the next job in the job queue. 

Although the inversion and the synthesis are currently made on a pixel-by-pixel basis, the code has been prepared for future extensions and the implementation of different methods, such as spatial coupling due to the telescope PSF \citep{VanNoort12} or sparsity constraints \citep{SparseInv}. These are incompatible with a pixel-by-pixel approach.

\subsubsection{Spatial regularization}
\label{spat_reg}

To invert a spatially extended field of view, we use a basic form of spatial regularization in order to discourage the appearance of salt-and-pepper noise in the inverted maps. This is essentially the application of various filters to the maps of node values after the inversion is performed, and then inverting the data again with the smoothed maps as initial guesses. This process is then repeated several times, until both the average $\chi^2$ and the maximum $\chi^2$ in the inverted field no longer changing significantly. We combine several different filters for spatial smoothing.
\begin{itemize}
 \item Median filter: this is the simplest filter used to avoid salt-and-pepper noise. This type of noise appears when the inversion method is
halted in a local minimum, or finds a solution that fits the data well, but departs from the solution in the neighboring pixels
because of the noise in the observations. 
 \item Gaussian filter: after the median filter, we apply a Gaussian filter to the map. As the general fit of the map improves, we decrease the width of the Gaussian.
 \item Wavelet filter: following \citet{SparseInv}, we are testing the applicability of wavelet compression to smooth the map. Between inversion cycles, we transform retrieved parameter maps to wavelet space, and keep only the low-frequency components. That is, we assume that the high-frequency components are only due to noise. This method effectively removes many artifacts, but if a departure from the spatially regularized solution is not caused by a halt in a local minimum, but is the result of a spectral feature or noise in the observations, the code truly finds the best fit, and regularization will not help.
\end{itemize}

\section{Tests}

The description of a method, or in this case, an inversion code, should include a number of (preferably repeatable) tests. We provide the results of three tests that should be easy to replicate by similar codes. They illustrate the applicability and the performance of SNAPI, the accuracy of the semi-analytical response functions, and an inversion of the Na\,I\,D lines.

\subsection{Retrieving the original model under noiseless conditions}

A level-zero test for a fitting procedure is its ability to recover the original model that was used to calculate a test dataset. If the model is analytical, derivatives are explicitly known, and a Levenberg-Marquardt minimization will converge to the minimum, given a reasonable initial guess. 

In the case of depth-dependent spectropolarimetic inversions, the problem is more complicated because the forward problem is not given analytically, but as a numerical solution of a differential equation (RTE), which in the NLTE case even requires an iterative approach to solve the strongly nonlinear and nonlocal coupling. Moreover, the derivatives are not analytically given, but computed either using a finite-difference approach, or as in the case of SNAPI, using a direct, semi-analytical method \citep{MMvN17}. If the derivatives are not accurate, the method will fail to converge to the original model, or worse, it might not converge at all. 


For this test, we used a transition in a simplified element, with a line strength $\log gf$ identical to that of H$\alpha$, and an elemental abundance identical to that of H. The element is identical to the first example from \citet{MMvN17}, and has a continuum, but only two energy levels, with Land\'{e} factors of one and zero for the upper and lower levels, respectively, so that the transition between them is a normal Zeeman triplet. This is an extreme NLTE example, since the line is strong and formed very high in the atmosphere. 

We started from a given set of atmospheric parameters (i.e., node values), synthesized Stokes profiles in the wavelength range from 6560 to 6567\,$\rm{\AA}$ with $10\,\rm{m\AA}$ sampling, and then inverted the spectrum, without adding any noise or instrumental effects. An ideal code should retrieve the original node values down to machine precision. We used a model with four temperature nodes, two nodes each in the magnetic field strength and in the
line-of-sight (LOS) velocity, and constant microturbulent velocity and magnetic field orientation. This is an atmosphere of a reasonable complexity to be diagnosed with one spectral line. The node values we used to generate the data and the initial guess solution for the fitting procedure are given in Table \ref{table2}.

\begin{table}
\caption{Node positions and the model values we used to generate the data and the initial model for the fitting procedure.} 
\label{table2}
\centering                                      
\begin{tabular}{c c c c}
\hline
Parameter & Node $\log \tau$ & True value & Initial value \\
\hline                                  
    $T\,[\rm{kK}]$ & \makecell{-4.8\\ -2.7,\\ -1.2\\ 0.0} & \makecell{5.9\\4.0\\5.2\\6.3} & \makecell{4.0\\5.0\\6.0\\7.0} \\     
\hline    $v_{\rm turb}\,[{\rm km/s}]$ & constant & 0.6 & 2.0 \\
\hline    $v_{\rm los}\,[{\rm km/s}]$ & \makecell{-5.0\\ -0.5} & \makecell{3.0\\0.0} & \makecell{-1.0\\1.0} \\
\hline    $B\,[{\rm kG}]$ & \makecell{-5.0\\ -0.5} & \makecell{1.7\\ 2.8} & \makecell{1.0\\ 1.0} \\
\hline    $\theta_{\rm{B}}\,[^o]$ & constant & 60.0 & 45.0 \\
\hline    $\phi_{\rm{B}}\,[^o]$ & constant & 30.0 & 10.0 \\
\hline                                            
\end{tabular}
\end{table}

Although the synthetic spectra in this example are noise-free, in the expression for $\chi^2$, a wavelength-independent noise equal to $10^{-3}$ of the continuum intensity was assumed for all four Stokes parameters to avoid dividing by zero. Since we
only minimized the $\chi^2$ and did not compare different models, the absolute level of the noise is not important, and we only provide it here in order to relate the values for $\chi^2$ shown below to the quality of the fit. Strictly speaking, the assumed level of noise would also influence the uncertainties, but since we did not apply noise here, and we retrieved the original model
exactly, and the uncertainties become meaningless. We used a starting value of $10^2$ for $\lambda$ in Eq.\,\ref{hessian} and increased or decreased it by a factor of 10 after successful
or unsuccessful correction attempts in the fitting procedure.

\begin{figure}
\includegraphics[width = 0.5\textwidth]{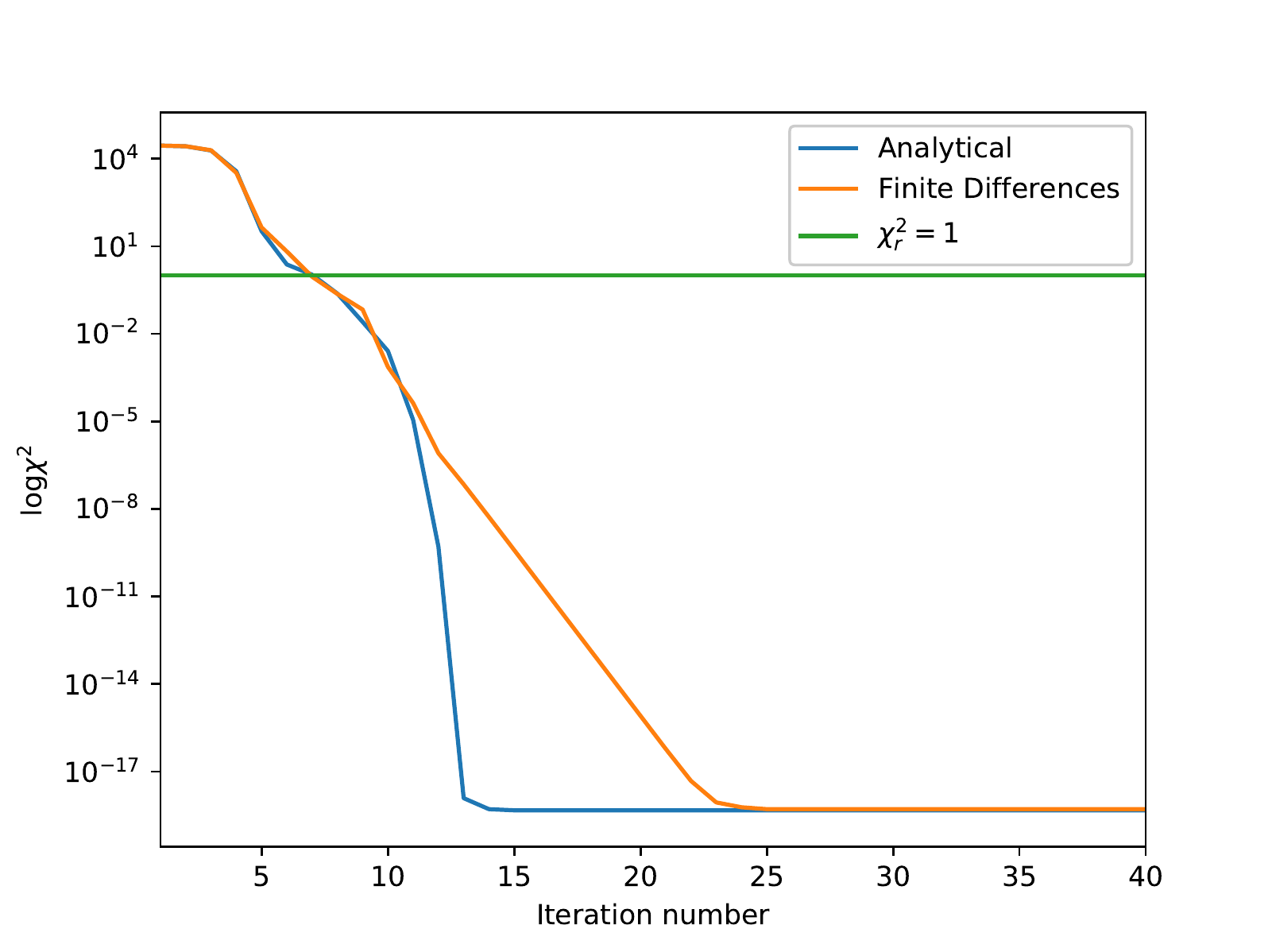}
\caption{Levenberg-Marquardt minimization using analytical and finite-difference response functions. For all practical purposes, the two methods converge equally fast in terms of iterations.}
\label{test_one_conv}
\end{figure}

Fig.\,\ref{test_one_conv} shows the convergence behavior (in terms of $\chi^2_{\rm r}$) of the Levenberg-Marquardt minimization using analytical and finite-difference derivatives. Here
\begin{equation}
\chi^2_{\rm r} = \frac{\chi^2}{N_{\lambda}-N_{\rm parameters}} 
\end{equation}
is the reduced $\chi^2$. As a rule of a thumb, when fitting data with a well-estimated amount of noise, a value of $\chi^2_{\rm r} \approx 1$ indicates a good fit. Lower values mean that the model overfits the data and is probably too complex, while higher values mean that the model underfits the data and is therefore not complex enough. This is strictly valid only for linear models \citep{chisq}, while for highly nonlinear ones, other methods should be used. However, we show the evolution of the reduced $\chi^2$ here as a simple way to compare the two methods we used to compute the response functions. 

Fig.\,\ref{test_one_conv} shows that both methods reach $\chi^2_{\rm r}\approx 1$ in same number of iterations. However, in this particular example, the analytical way of computing the response functions is a factor of 3 faster, which translates into a factor of 3.
This clearly saves computing time. Both methods reach $\chi^2_{\rm r}\approx 0$ (we note that there is no noise in the data), which translates into a retrieval of the original model down to eight digits. Additionally, the analytical method converges faster once very high agreement is achieved. The reason might be a sub-optimal choice of the step size in the finite-difference calculations. Even if this were the case, it just illustrates another advantage of the analytical approach: there is no need to fine-tune any step size. We have repeated this test on several different atmospheric models and found practically identical results. We conclude that the analytical response functions are accurate enough and that our inversion procedure converges well on ideal, noise-free data.

\subsection{Fitting synthetic data with noise}

We then tested the code on a more realistic example, again using synthetic data. We considered a five-level model of the Ca\,II atom and focus on the spectral line at $8542$\,$\rm{\AA}$, one of the most frequently used NLTE spectral lines for chromospheric diagnostics \citep[e.g.,][]{HectorI,Ca_diag_Jaime}. The spectrum is synthesized from one vertical column of the BIFROST enhanced network simulation, shown in Fig.\,\ref{test2}, described in \citet{BIFROST}. Following \citet{Flashinversion}, we added a depth-independent microturbulent velocity of 3\,km/s for a greater
similarity of the line shape to the observed ones. Although we added Gaussian wavelength-independent noise ($1\times10^{-3}\,I_{\rm c}$) to all Stokes parameters, which is sufficient to overwhelm the Stokes $Q$ and $U$ signals, we still fit them, in order to have better constraints for the inclination. To invert the data, we used a model with seven nodes in temperature, five in the LOS velocity, and two in the magnetic field strength. The microturbulent velocity and the orientation of the magnetic field were assumed to be constant throughout the atmosphere (but were still free to vary).  We inverted the synthetic data using the analytical response functions multiple times, starting from different initial guesses, and chose the model with the lowest $\chi^2$. We then used the finite-difference response functions, starting from the best initialization, and compared the convergence properties, as in the previous example.

\begin{figure}
\includegraphics[width = 0.5\textwidth]{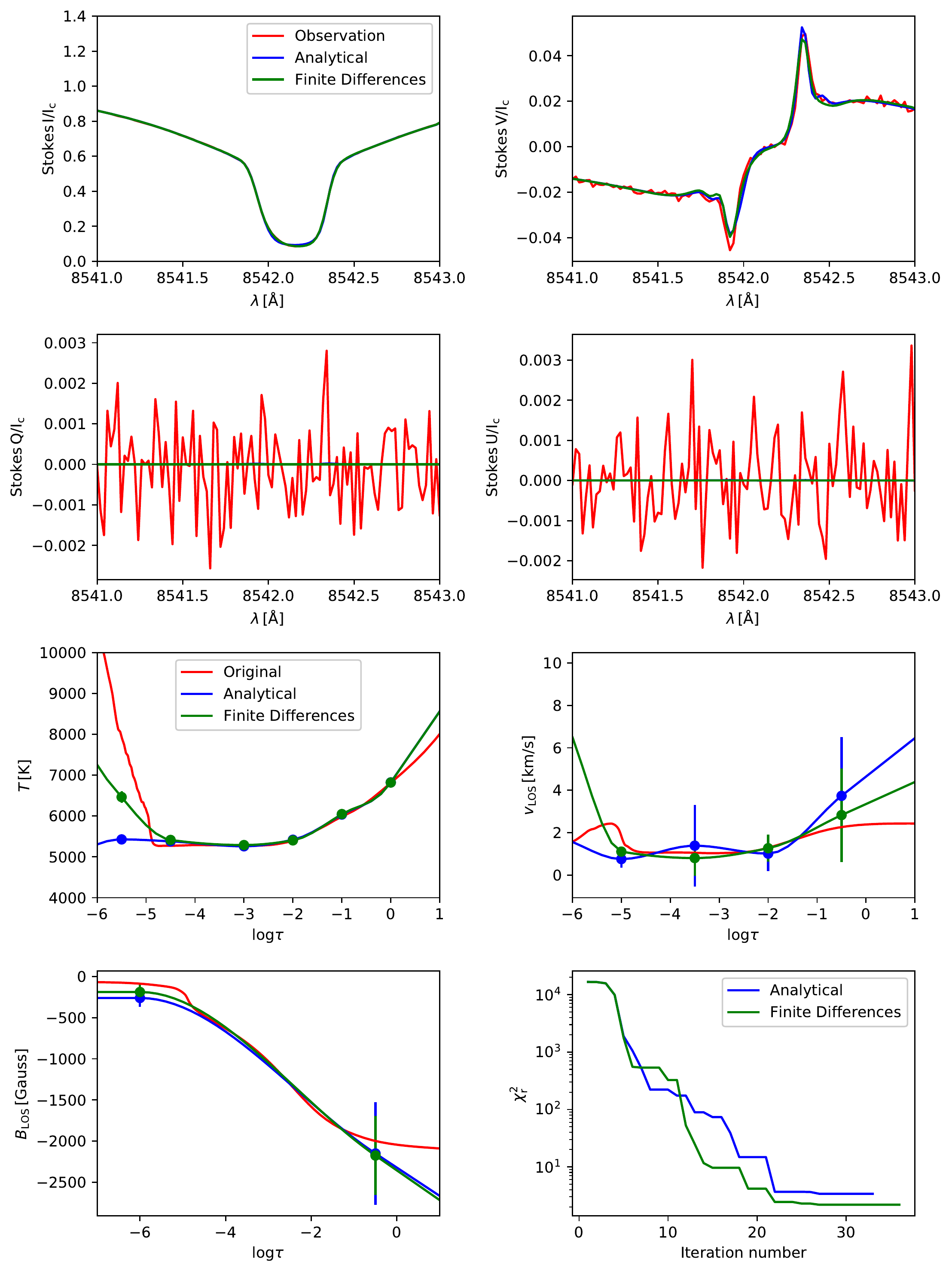}
\caption{Comparison of an inversion of noisy synthetic data using SNAPI, relying on analytical and finite-difference response functions. The uppermost four panels show the agreement between the original and the fitted spectra. The next three panels show the inferred stratifications of temperature, LOS velocity, and LOS magnetic field and the comparison with the original stratification from the MHD cube. The error bars are plotted three times larger for
clarity. Finally, the lower right panel shows the convergence comparison between inversions that use analytical and finite-differences response functions.}
\label{test2}
\end{figure}

Fig.\,\ref{test2} shows the agreement between the synthetic observations and the best fits obtained using the SNAPI inversion code. The analytical and finite-difference response functions both result in an excellent fit to Stokes\,$I$ and a very good fit to Stokes\,$V$, while the remaining two Stokes parameters are below the noise. The fit quality of $V/I_{\mathrm c}$ would be better if we had
given more freedom to the magnetic field or used different weighting, but the aim of this test was to verify the performance of the analytical response functions, therefore we postpone the discussion on choices for atmospheric parameterization to a subsequent work. The inferred atmospheric stratifications agree well with the original MHD atmosphere. More careful node placement and fine-tuning of the model would result in even better agreement, but this tuning strongly depends on the lines and observations, therefore we do not discuss Ca\,II inversions more deeply here, and we refer to \citet{Ca_diag_Jaime,Ca_CQ} and \citet{Flashinversion},
for instance, for more details. Finally, we show that the convergence of the two methods for calculating the response functions is quite similar, as was the case in the previous example. In this case, the finite-difference approach converges slightly faster, resulting in a difference of several iterations. The computing time difference, however, is significant, since finite-difference
based response functions are more than a factor of 10 slower in this case because of the more complicated atomic model and the larger number of nodes used for the atmospheric structure. This examples illustrates the significant time savings well that can be obtained with analytical response functions.

In Fig.\,\ref{test2} we also show the agreement between original and inferred stratification of the temperature, velocity, and magnetic field, as well as the estimates of the uncertainties. All the parameters are constrained fairly well with our model, and models obtained with analytical and numerical response functions agree with each other. The error bars are just guidelines that provide us with a rough estimate of how well the parameters are constrained, but probably do not represent reliable uncertainties.

\subsection{Inverting synthetic data from an MHD cube}
A typical example of NLTE lines are the Na\,D lines at 5889 and 5896\,\AA. Although the opacity of the line and an analysis of the contribution function suggest that these lines are formed in the chromosphere because the formation of these lines is dominated by scattering, they are sensitive to the temperature much deeper down \citep[for an in-depth discussion, see][]{Na_Bruls,Na_Jorrit}. 

Like the Ca\,II infrared line from the previous example, the Na\,D lines are NLTE lines, formed above the photosphere, and are only moderately sensitive to the magnetic field (the Land\'{e} factors of the D1 and D2 lines are 1.33 and 1.17, respectively). Unlike the Ca\,II infrared line, however, they have not been extensively used in spectropolarimetric inversions, and the diagnostic potential is therefore largely unknown.

As a first step toward improving this situation, the spectrum containing the Na\,I\,D lines, as well as two LTE photospheric lines in between, which are magnetically sensitive, and strong enough to provide some additional diagnostic potential, were synthesized for a small patch ($40\times40$ pixels) of a BIFROST enhanced network simulation (\citet{BIFROST}. The main properties of all four lines considered for this example are summarized in Table\,\ref{table_ex_3}.

\begin{table}
\caption{Wavelengths, Land\'{e} factors, and formation (sensitivity) regions of the four lines we used to calculate and invert the
synthetic spectra} 
\label{table_ex_3}
\centering                                      
\begin{tabular}{c c c c}
\hline
Line &  $\lambda\,[\AA]$ & $g_{\rm L}$ & Region \\
\hline                                  
    Na\,I\,,D2 & 5889.9 & 1.17 & Photosphere/Chromosphere \\
    Fe\,I & 5892.7 & 1.83333 & Photosphere \\
    Ni\,I & 5892.9 & 1.0 & Photosphere \\
    Na\,I\,D1 & 5895.9 & 1.33 & Photosphere/Chromosphere \\
\hline                                            
\end{tabular}
\end{table}

The main goal of this test is to show the inversion results of a realistic artificial dataset, and evaluate the agreement between the original atmospheres, generated by an MHD code, and the atmospheres inferred by SNAPI. We synthesized the spectral region in the range from 5886 to 5896, with $0.01\,\rm{\AA}$ sampling, in a patch of the MHD cube that contains a strong magnetic field. The data were spectrally degraded assuming a Gaussian broadening function with an FWHM of $30\,\rm{m\AA}$. We then added wavelength-dependent Gaussian noise to the synthetic observations, assuming a signal-to-noise ratio level of $10^3$, and inverted the observed Stokes $I$ and $V,$ since, similarly to the previous example, the linear polarization signals were well below the noise level. We used an atmospheric model with four temperature nodes, placed at $\log\tau=-2.5,-1.4,-0.6,0.0$, two nodes in the LOS velocity, placed at $\log\tau=-3.3,-0.5,$ and two nodes in the magnetic field ($\log\tau=-2.9,-0.5$). 

We have attempted many different node combinations for this inversion, and while increasing the number of nodes in the magnetic field and LOS velocity does yield better fits, the inferred atmospheres show unrealistically strong depth variations of these parameters. These variations can result in unphysical results, such as failure to conserve the magnetic flux over the inverted region. 

The inversions for this test were carried out using SNAPI with analytical response functions, and used the multi-cycle spatial regularization procedure described in Section\,\ref{spat_reg}. The spatial regularization consisted of 10 ``regularization cycles'', alternated with bursts of 15 Levenberg-Marquardt iterations each. In each cycle, except for the last, median and Gaussian filters were applied to the intermediate results, and the width of the Gaussian filter was gradually decreased from 5 to 1 pixel. 

\begin{figure*}
\includegraphics[width = \textwidth]{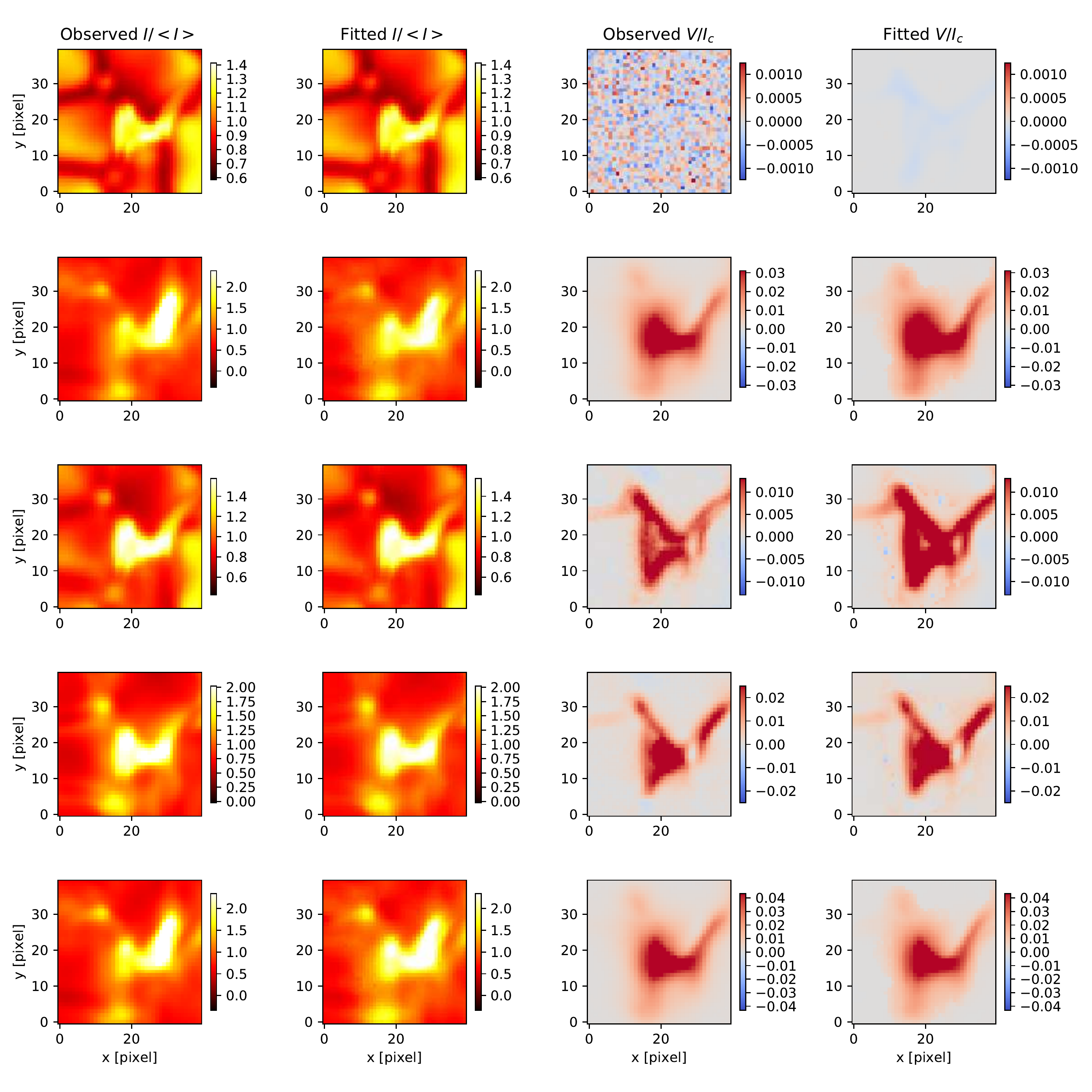}
\caption{Comparison between synthetic observations (first and third column) and the best-fit solutions (second and fourth column) at five different wavelengths. Top to bottom: Continuum, Na\,D2, Fe, Ni, and Na\,D1 line. The intensity is plotted at the line center wavelength, and Stokes\,$V$ is averaged over $50\,\rm{m\AA}$ from the line center toward the red.}
\label{test3}
\end{figure*}

In Fig.\,\ref{test3} we compare the synthetic and inverted observations at the continuum and line center wavelengths of the four lines from Table \ref{table_ex_3}. Qualitatively, the agrement in Stokes $I$ is very good, and there is almost no trace of salt-and-pepper noise in the fitted spectra. This plot nicely illustrates that the spectra look different at these five wavelengths, since they encode information from different depths in the atmospheres. The agreement between synthetic and fitted Stokes $V$ is not as good. The suspected main reason for this is the simplicity of the atmospheric parameterization, which is unable to accurately reproduce the spectra from pixels on the border between the magnetized and non-magnetized regions. Presumably, the magnetic field strength there varies strongly with depth, and the assumption of a linear height dependence is not accurate. 

\begin{figure}
\includegraphics[width = 0.5\textwidth]{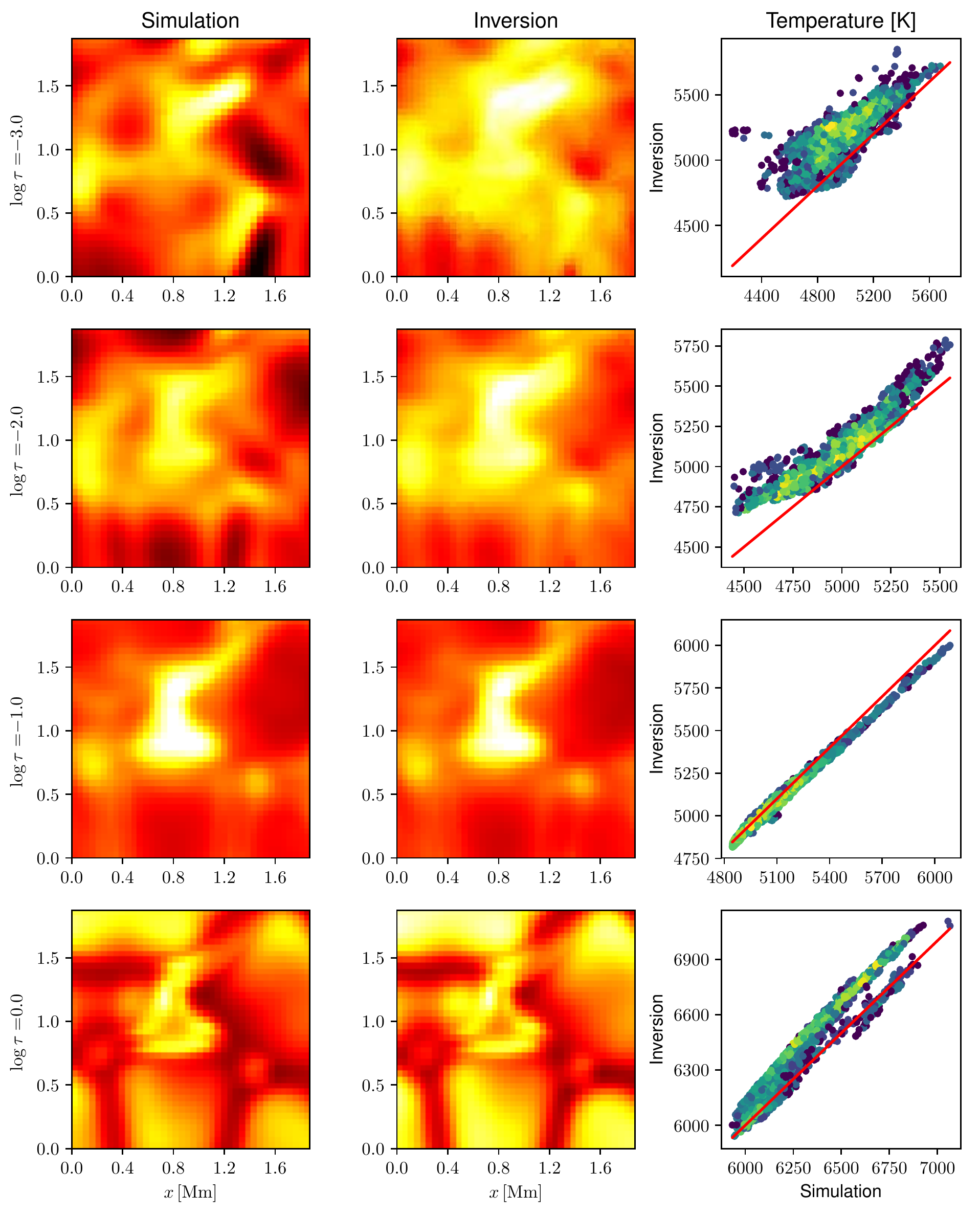}
\caption{Comparison between inferred (left column) and original temperature (right column) at three different optical depths. To facilitate the comparison, we also include scatter plots (right column).}
\label{test3_T}
\end{figure}

Good agreement of the observables, however, does not guarantee that our inferred atmospheric structure is also accurate. Fortunately, since the spectra were synthesized from an MHD atmosphere, the inversion results can be directly compared to the true atmosphere. In Fig.\,\ref{test3_T}, we compare the temperature stratification between the fitted atmospheres and the original BIFROST cube. While the temperature at the two lowest depth points ($\log \tau = -1$ and $\log \tau = 0$) is almost identical, agreement in the upper layers is not so good. This is not surprising: although the Na\,D lines are very opaque, they are practically insensitive to the temperature in the upper layers as a result of their scattering nature, as has been stressed some decades ago by \citet{Na_Bruls}. The addition of the photospheric lines of Fe and Ni are not especially helpful here, as they are too weak to be sensitive to higher atmospheric layers. 

Additionally, the node positions may play a role, since placing the nodes in the upper layers is necessary to avoid large temperature gradients, but it might very well be that the higher temperature nodes are largely degenerate among themselves or with respect to other parameters. Clearly, this spectral region is not ideal for temperature diagnostics.

\begin{figure*}
\includegraphics[width = 0.5\textwidth]{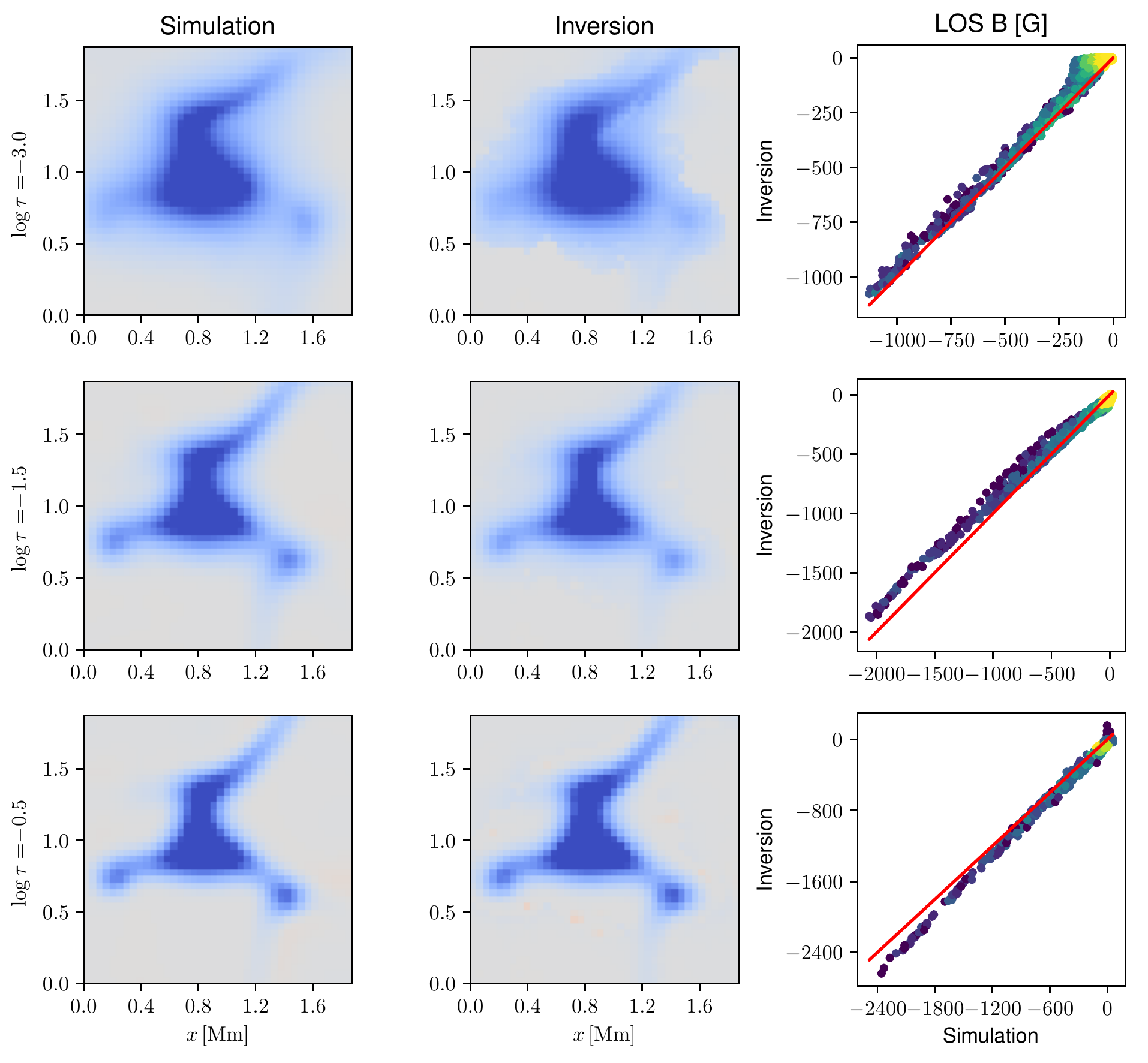}
\includegraphics[width = 0.5\textwidth]{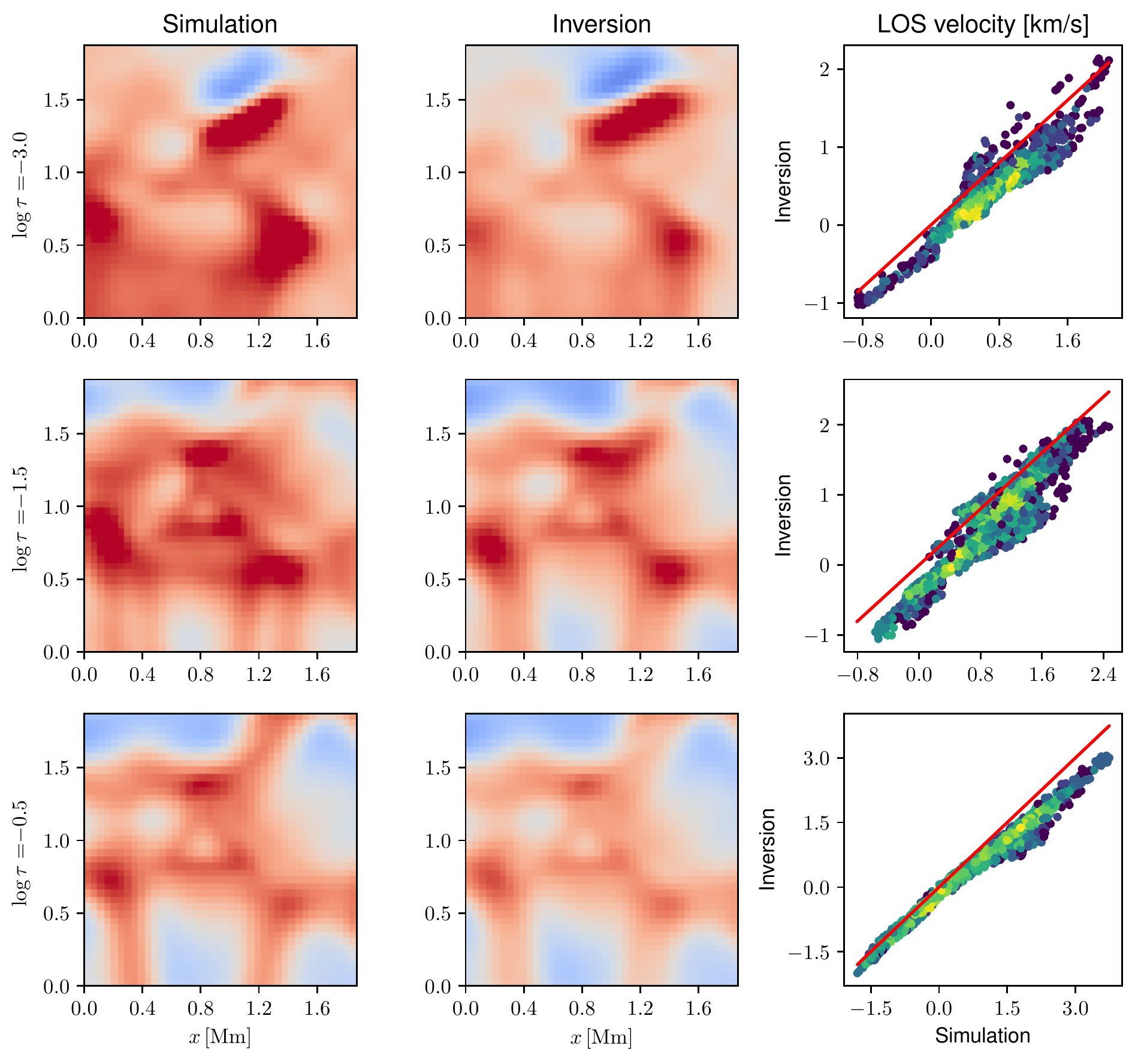}
\caption{Same as Fig. \ref{test3_T}, but for the LOS magnetic field (leftmost three columns) and the LOS velocity (rightmost three columns).}
\label{test3_BV}
\end{figure*}

Fig.\,\ref{test3_BV} shows the same comparison, but for the LOS velocity and magnetic field. In contrast to the temperature, the
agreement here is very good throughout the whole range of optical depths from $\log\tau=-3$ to $\log\tau=-0.5$. The reason is that the magnetic field and the velocity influence the line profile mostly locally, and this influence is indifferent to whether the line is formed by scattering or by pure thermal emission. There is still some influence of the velocity and the magnetic field on the level populations, but this is only important in the case of large velocity gradients and very strong magnetic fields. The spatial distribution of the velocity in the $(x,y)$ plane seems to change significantly with optical depth, and interestingly enough, we are able to capture the variations using only two velocity nodes. Variations in the magnetic field are much simpler: the magnetic field weakens and diffuses, which suggests a canopy-like structure. 

Although the two-node model captures the behavior of the magnetic field in the deeper layers quite well, it appears to create a halo around the magnetic structure in the center of the FOV, in the upper atmospheric layers (see the top leftmost panel of Fig.\,\ref{test3_BV}). This halo is the reason for the reduced level of agreement between the synthetic observations and best fits, seen in Stokes $V$ in Fig.\,\ref{test3}. An attempt to remedy this by adding a node in the magnetic field strength resulted in a halo of pixels where the variation of the magnetic field with height is non-monotonic. Although the fits are better, the inferred atmospheric stratification is worse.

\section{Discussion and conclusions}
We have developed and described a spectropolarimetric inversion code that is able to invert LTE and NLTE lines. The code is called SNAPI, which is short for spectropolarimetric NLTE analytically powered inversion, and it is based on accelerated lambda iteration, as described in \citet{RH1}, and uses the analytic approach for computing the response functions \citep{MMvN17}. This results in a significantly improved performance over codes using finite differences. To fit a model atmosphere to the observed spectrum, it uses a simplified atmospheric model based on nodes, where the parameters are assumed to vary between them according to a specific interpolation scheme (in our case, a second-order Bezier interpolation scheme).

To illustrate the potential applications of SNAPI and to establish a framework for comparison with similar codes, we showed the results of three tests. Using an academic two-level atom, we synthetized the spectrum using a given, node-based, atmospheric structure, and inverted the synthetic observation. The results showed the accuracy of the analytical response functions, and confirmed that the code is capable of finding the correct solution to a very high precision, under ideal conditions. In the second test, using a more realistic atmosphere from an MHD simulation, we synthesized the profile of the Ca\,II 8542 line, added noise, and inverted the synthetic observation. In the final test, we inverted a synthetic observation of the spectral region around the Na\,D lines, calculated from a patch of a publicly available enhanced network simulation \citep{BIFROST}. 

In all three tests, the code returned good fits to the observed data and inferred atmospheres that agree reasonably well with the ``true'' ones. In the final test, however, we found that a better fit to the data does not necessarily guarantee that the solution is better as well. An inappropriate model can result in an unphysical or unnecessary complicated solution that is not a good description of the true solution at all. 

With the current, node-based, regularization strategy, it is entirely up to the inverter, and not the inversion code, to determine the optimal regularization setup, and make estimates of the corresponding errors in the results. It is more than likely that the choice of nodes and fitting weights has a much greater influence on the inversion results than the choice of code, the minimization procedure, or the noise level. This means that even though the inversion itself is an powerful tool for spectropolarimetric diagnostics, careful and insightful interpretation of the results, combined with a good understanding of the inherent limitations of the methods used to obtain them, will remain essential.


\begin{acknowledgements}
We thank Smitha Narayanamurthy and Rafael Manso Sainz for their invaluable comments on the manuscript and Francisco Iglesias, Sami K. Solanki, Andr\'{e}s Asensio Ramos and Jaime de la Cruz Rodriguez for many stimulating discussions and suggestions during the development of the code.
 
This research is made using the supercomputing capabilities at the Max Planck Institute of Solar System research and the Gesellschaft
f\"{u}er wissenschaftliche Datenverwertung in G\"{o}ttingen (GWDG). Plots and post-processing of the data were made using python with the matplotlib, numpy, and scipy packages. The manuscript has made use of the NASA astrophysics data system.
\end{acknowledgements}

\bibliographystyle{aa}  
\bibliography{inversion} 

\end{document}